\documentclass[aps,pra,reprint,amssymb,superscriptaddress,nofootinbib,floatfix]{revtex4-2}
\usepackage{booktabs}
\usepackage[strings]{underscore}
\usepackage{graphicx}
\usepackage{amsmath,amssymb,bm}
\usepackage{braket}
\usepackage{xcolor}
\usepackage{tikz}
\usepackage{ragged2e}
\usepackage[most]{tcolorbox}
\tcbuselibrary{breakable,skins}
\usetikzlibrary{positioning,arrows.meta,shapes.geometric,calc,fit,backgrounds}

\definecolor{iqonblue}{HTML}{2E3B5E}
\definecolor{iqonorange}{HTML}{8B5A2B}
\definecolor{iqongreen}{HTML}{4F6470}
\definecolor{iqongray}{HTML}{606060}
\definecolor{linkdark}{HTML}{1F3A82}

\usepackage[colorlinks=true,citecolor=blue,linkcolor=red,urlcolor=blue]{hyperref}

\newtcolorbox{promptbox}[1][]{
    colback=gray!5,       
    colframe=gray!60,     
    leftrule=3pt,         
    rightrule=0pt,        
    toprule=0pt,
    bottomrule=0pt,
    sharp corners,         
    fonttitle=\bfseries,
    coltitle=black,
    attach title to upper,
    after title={\smallskip\newline},
    before upper={\small\ttfamily\justifying},
    boxsep=5pt,
    arc=0mm,
    breakable,
    #1
}

\usepackage{orcidlink}

\usepackage{afterpage}

\begin{document}

\title{Lowering the implementation barrier of neutral-atom quantum computing with agentic workflows}

\author{Constantin Dalyac~\orcidlink{0000-0002-0339-6421}}
\author{Alexandre Dauphin~\orcidlink{0000-0003-4996-2561}}
\author{Loïc Henriet~\orcidlink{0000-0003-3108-0595}}
\affiliation{PASQAL SAS, 24 rue Emile Baudot, 91120 Palaiseau, France}

\author{Christophe Jurczak~\orcidlink{0009-0001-8204-7333}}
\affiliation{Quantonation Ventures, Dallas, TX 75206, USA}

\date{July 28, 2026}
\begin{abstract}
    Quantum computers are moving from research laboratories to industrial machines accessible via the cloud and integrated into high-performance computing facilities. However, translating theoretical quantum protocols into hardware experiments remains a major bottleneck, requiring expertise across protocol design, compilation, simulation, and cloud execution. Here, we introduce an agentic workflow that automates this pipeline on neutral-atom quantum processors (here two Pasqal QPUs available on the cloud) while keeping the researcher in the loop for critical validation. In three case studies from many-body physics and optimization, the agent went from published paper or patent to a QPU campaign run overnight. In particular, human intervention was crucial to ensure scientific validity: the agent selected an inadequate observable in one experiment and constructed a plausible but incorrect hardware diagnosis in another, with both failures detected only through domain-expert review. Finally, we use a second agent to classify a corpus of 633 Rydberg-array arXiv papers and show that nearly half are implementable on present-day QPUs while identifying specific hardware upgrades needed for the rest. Together, these results demonstrate that agentic workflows provide a practical bridge between theoretical ideas and physical hardware, opening quantum experimentation to a much broader scientific community.
\end{abstract}

\maketitle

\begin{figure*}
    \centering
    \includegraphics[width=0.9\linewidth]{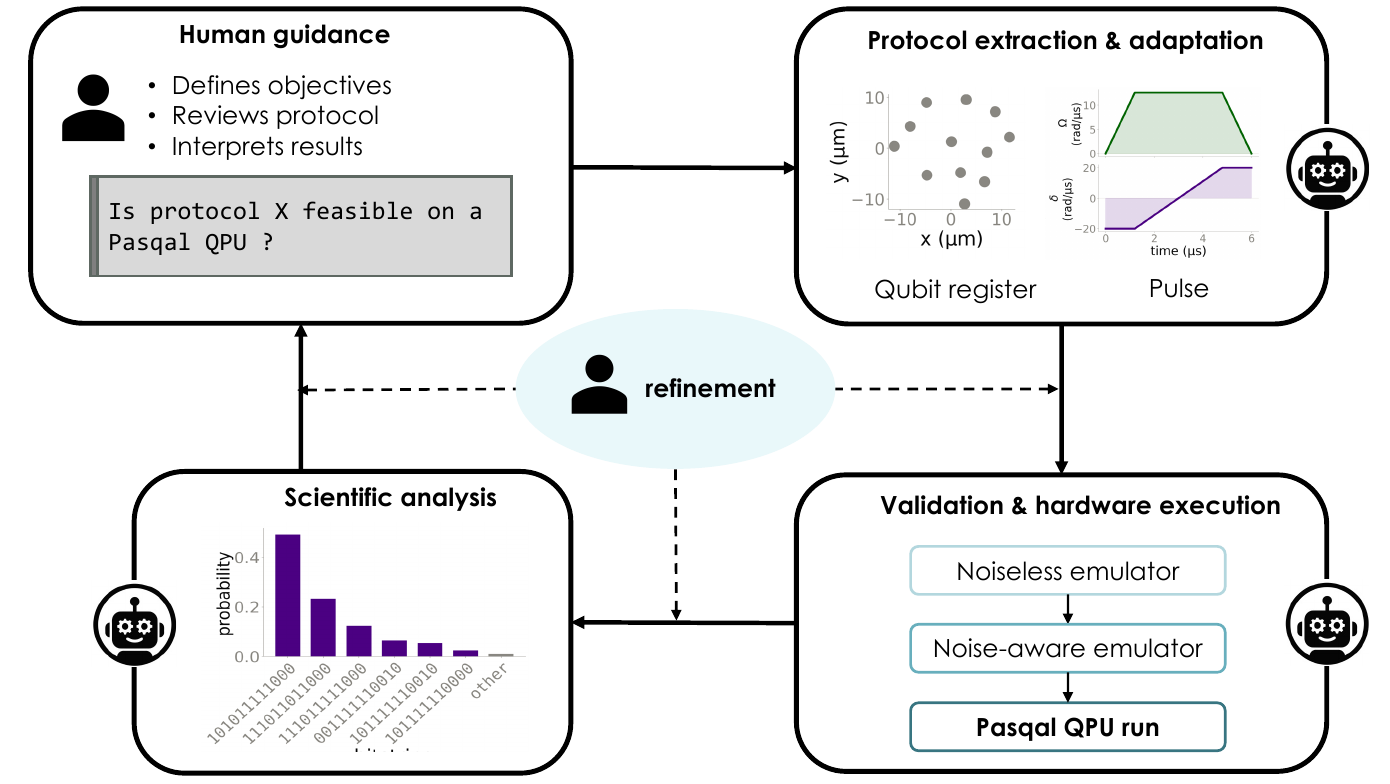}
\caption{\textbf{AI-assisted workflow for translating quantum protocols into hardware experiments}. A researcher defines the scientific objective and interacts with the AI workflow through natural-language instructions. The AI extracts or generates an experimental protocol, including the neutral-atom register and pulse sequence, from a scientific publication, patent, or user request. The pulse sequence is determined by the qubit register, the time evolution of the Rabi frequency $\Omega$ (rad.$\mu s^{-1})$ and the detuning $\delta$ (rad.$\mu s^{-1})$. Before execution on quantum hardware, the protocol is validated using noiseless and noise-aware emulation to identify implementation issues and support iterative refinement. Once validated, the experiment is executed on a cloud-accessible neutral-atom quantum processor. The resulting measurement data are automatically processed into scientific observables, which the researcher interprets to assess the outcome and, when necessary, refine the protocol for subsequent iterations. The workflow keeps the researcher in the decision loop while automating the engineering tasks required to bridge the gap between a scientific idea and a hardware experiment.
}
    \label{fig:workflow}
\end{figure*}

\section{Introduction}
Validating a quantum-computing application ultimately means confronting a theoretical proposal with physical hardware.
For much of the field's history, the binding constraint on that step was hardware availability: quantum processors were scarce, laboratory-bound systems accessible mainly to the groups that built them. 
Today, however, quantum processing units (QPUs) are increasingly stable, remotely accessible resources embedded in cloud and high-performance-computing ecosystems.
Neutral-atom platforms are a representative case: arrays of individually controlled atoms interacting through
Rydberg states realize programmable many-body
Hamiltonians~\cite{browaeys2020manybody}, and scale naturally into the 100--1000-qubit range with flexible geometries and both analog and digital modes of operation~\cite{henriet2020neutralatoms,menssen2026strategic}.
This versatility supports applications spanning quantum simulation, graph machine learning and combinatorial optimization~\cite{ebadi2022mis, dalyac2024graph}.

Accessible hardware, however, does not imply easy experimentation. Between a published protocol and a QPU lies a long chain of research and engineering tasks: extracting the procedure from the literature, translating it into device-level controls, validating it in a simulation, navigating cloud-execution infrastructure, and interpreting noisy data. Each step is individually well understood and the ensemble is well supported by a software stack, but the full chain demands fluency across quantum physics, software engineering, numerical simulation, cloud computing, and hardware characterization. The resulting \emph{implementation barrier} 
often exceeds the effort of formulating the original scientific idea, leaving many physically viable proposals unexplored despite suitable hardware being available.

This barrier is, at its core, a coordination problem across specialized domains, precisely the kind of problem that recent agents powered by large language models (LLM) have begun to address. Agentic systems can orchestrate heterogeneous software tools, scientific workflows, and laboratory infrastructure~\cite{boiko2023coscientist,bran2023chemcrow,lu2024aiscientist}, carrying a task across interfaces that have traditionally required separate human specialists.
Rather than replacing scientific judgment, such systems offer a way to fold expertise distributed across technical layers into a single operational workflow.

Building on this observation, our contributions are threefold. First, we introduce an agentic workflow that carries a quantum-computing task all the way from a prompt to hardware execution. The workflow is tailored for Pasqal's QPUs, and executes specifically on FC1 and SA1, which are located respectively in Canada and in Saudi Arabia.  Second, we validate the framework across three increasingly difficult applications: two many-body quantum simulations and a graph-coloring algorithm. Finally, we apply a separate, lighter agentic workflow to the literature itself: it screens arXiv for Rydberg-array theory papers compatible with the platform, using back-of-the-envelope rather than detailed calculations. Of the 526 papers classified, roughly half are viable starting points for experimentation on present-day hardware. For the remaining papers, the workflow identifies the hardware specifications that would need to be reached, providing direct feedback for the design of future hardware.

\section{Agentic workflow}\label{sec:wf}

We start by describing known agentic workflows that have been developed for scientific research, before presenting our specific workflow for this study, presented at a high-level in Fig.~\ref{fig:workflow}. 

\subsection{Prior work}
LLM agents are now used to run increasingly autonomous scientific workflows, in which they coordinate tools, execute multi-step procedures, and interface with external computational or experimental infrastructure~\cite{zheng2025automation}.

Coscientist~\cite{boiko2023coscientist}, ChemCrow~\cite{bran2023chemcrow}, and the AI Scientist~\cite{lu2024aiscientist} established the feasibility of agentic automation: the first two translate natural-language objectives into executable chemistry experiments through tool-augmented reasoning, while the AI Scientist automates machine-learning from a hypothesis to a manuscript.
Role-specialised multi-agent systems (AtomAgents~\cite{ghafarollahi2025atomagents}, ProtAgents~\cite{ghafarollahi2024protagents}) outperform monolithic ones on design tasks, and self-driving laboratories~\cite{tom2024sdl} run modular instruments in closed loops (e.g.\ ChemOS~2.0~\cite{sim2024chemos}, with multi-agent benchmarks~\cite{mandal2025afmbench}).

Within quantum technologies, reinforcement-learning controllers handle real-time qubit feedback~\cite{reuer2023realtime} and error-correction stabilisation~\cite{sivak2025rlqec}, and LLM-based program synthesis (Agent-Q~\cite{jern2025agentq}, with retrieval~\cite{asif2025pennylang} or simulator feedback~\cite{mikuriya2025qcoder}) generates circuits from specifications. Multi-agent architectures have similarly been applied to translate natural-language requirements into traceable quantum-application workflows~\cite{tao2026qpipe}. Most closely, Shiraishi~\emph{et al.}~\cite{shiraishi2026mcp} expose QPU and HPC execution to an agent through a Model Context Protocol server. These approaches lower the expertise to operate hardware but assume a well-defined target. AI-Mandel~\cite{arlt2025aimandel} goes one step further and lets the agent choose its own research question in quantum optics, using a domain-specific design tool to
turn each idea into an implementable experiment.

As agentic workflows become autonomous, they also raise questions about the evolving role of the human scientist. Consequently, many recent frameworks emphasize the importance of maintaining meaningful human oversight throughout the research process~\cite{kumar2024applications, tang2025risks}. Systematic evaluations have shown that LLMs can generate persuasive yet fundamentally incorrect analyses due to hallucinations and flawed reasoning~\cite{messeri2024illusions,huang2024selfcorrect}. Recent benchmarks on open physics problems further demonstrate that even advanced autonomous agents struggle with the tacit domain knowledge, contextual judgment, and physical intuition required for scientific discovery~\cite{malinowski2026aiphysics}. We observe similar limitations in this study, particularly during result interpretation, and discuss these challenges throughout the case studies presented in this work.

\subsection{Description of our agentic workflow}

Our framework builds on Pasqal's existing software stack: Pulser~\cite{silverio2022pulser}, an open-source Python library for designing pulse sequences on neutral-atom QPUs and their emulators; the matrix-product-state emulator emu-mps~\cite{bidzhiev2025emulation}; and the Pasqal Cloud SDK~\cite{pasqal-cloud-sdk-github,wennersteen2026hybrid}, which provides programmatic access to the QPUs, their calibration data, and noise models.

On top of this stack, the workflow is implemented as a collection of skills, each responsible for one stage such as: translating a scientific source into a specification, compiling that specification into a pulse sequence, validating it in emulation, executing it on hardware, or analyzing the returned data (Fig.~\ref{fig:workflow}). The individual skills are detailed in Appendix~\ref{app:workflow}. We here use Anthropic’s Claude Code with the models Opus 4.8 and Fable 5.

The skills operate on a single structured artifact, a machine-readable experimental specification $\texttt{experiment_spec.json}$ capturing the physical parameters, observables, and constraints of the experiment. This specification is the primary interface between researcher and agent: it can be inspected, modified or approved before anything is executed. By construction, any modification propagates to all downstream stages.

Importantly, no protocol reaches the QPU without passing through emulation. A candidate sequence is first simulated noiselessly to establish a reference prediction, then re-simulated with device-specific noise parameters retrieved from the cloud SDK. Comparing the two indicates whether the target observable is expected to survive realistic hardware conditions and helps the researcher in deciding to submit.

Finally, the skills can be composed into iterative refinement loops that follow the researcher–AI dialogue: validation or hardware results may motivate changes to the register geometry, pulse schedule, or target observable, upon which the agent regenerates the sequence and re-runs the relevant stages.

\section{End-to-end examples}\label{sec:ex}

We present three examples of increasing complexity. First, we reproduce a well-known quantum simulation result expected to be achievable on FC1 and SA1 (Example~A). Next, we examine a theoretical proposal to assess whether the predicted phase of matter can be realized on FC1 (Example~B). Finally, we address an optimization application derived from a patent, specifically a graph-coloring algorithm (Example~C).

Before turning to the examples, we briefly recall the control knobs these experiments are expressed in. A neutral-atom QPU realizes an Ising-like Hamiltonian whose terms map onto three handles: the \emph{qubit register}, i.e.\ the positions of the atoms, which fix the pairwise interactions $C_6/r_{ij}^6$ through the chosen geometry; the \emph{Rabi frequency} $\Omega(t)$, which coherently couples the ground and Rydberg states and sets the transverse field; and the \emph{detuning} $\delta(t)$ of the driving laser from resonance, which acts as a longitudinal field controlling the Rydberg population. A protocol is thus a register plus time profiles $\bigl(\Omega(t),\delta(t)\bigr)$, typically applied globally to all atoms.
Each handle comes with limits: a maximum number of atoms and a minimum interatomic spacing for the register, caps on $\Omega$ and on the reachable detuning range, finite ramp rates, and a maximum sequence duration. These bounds are device-specific and evolve with hardware upgrades. 

\subsection{Reproduction of a literature hardware result}\label{sec:ex_keesling}

\subsubsection{Context and task}

Our objective is to reproduce known results that have already been run on hardware. We select a landmark paper by Keesling \textit{et al.}~\cite{keesling2019quantum} from 2019 where the authors study the Kibble-Zurek mechanism when driving a 1D array of atoms through a quantum phase transition from the paramagnetic to several ordered states, namely $\mathbb{Z}_2, \mathbb{Z}_3$ and $\mathbb{Z}_4$. From a domain expert point of view, we know that the quantum phase transition to $\mathbb{Z}_2$ is realizable but not the higher order ones. 

We intentionally prompt the agent towards the $\mathbb{Z}_3$ phase to judge if the workflow can recognize hardware-imposed limitations. We use the following prompt:

\begin{promptbox}[title={Agent Prompt}]
Reproduce the Z3 density-wave results from Keesling et al. on FC1. Read the paper, identify the protocol for the Z3 phase, adapt it to FC1 constraints, and run the pipeline through EMU validation. If the EMU verdict is GO, propose a QPU submission plan. If any aspect of the paper's protocol cannot be faithfully reproduced on FC1, document the deviation and justify whether it preserves the essential physics.
\end{promptbox}

\subsubsection{Workflow and dialogue}

The agent extracts the protocol directly from the paper: a 51-atom 1D chain, the
blockade ratio $R_b/a$ defining the ordered lobe ($R_b/a \approx 2.16$--$2.43$ for
$\mathbb{Z}_3$, with $R_b$ the blockade radius and $a$ the atomic spacing), and the pulse schedule together with the sweep rates tabulated for each interaction range. It then fetches FC1's live device specification through the cloud SDK and combines the two to reach the following verdict:

\begin{itemize}
    \item[\textit{(i)}] The $\mathbb{Z}_3$ phase is not native: reaching the targeted regime at the spacing floor forces $\Omega/2\pi$ down to 0.087\,MHz, stretching the dynamical timescale so that neither one Rabi period nor the nearest-neighbour interaction time fit inside the sequence cap. By the same scaling argument, the still-higher $\mathbb{Z}_4$ phase places it further out of reach.
    \item[\textit{(ii)}] The $\mathbb{Z}_2$ phase on the contrary is native: at the 5\,$\mu$m spacing floor and full $\Omega$, $R_b/a = 1.28$ falls well inside the ordering lobe, and the paper's slowest schedule fits the 6\,$\mu$s FC1 specification.
\end{itemize}

\begin{figure}[b]
    \centering
    \includegraphics[width=\linewidth]{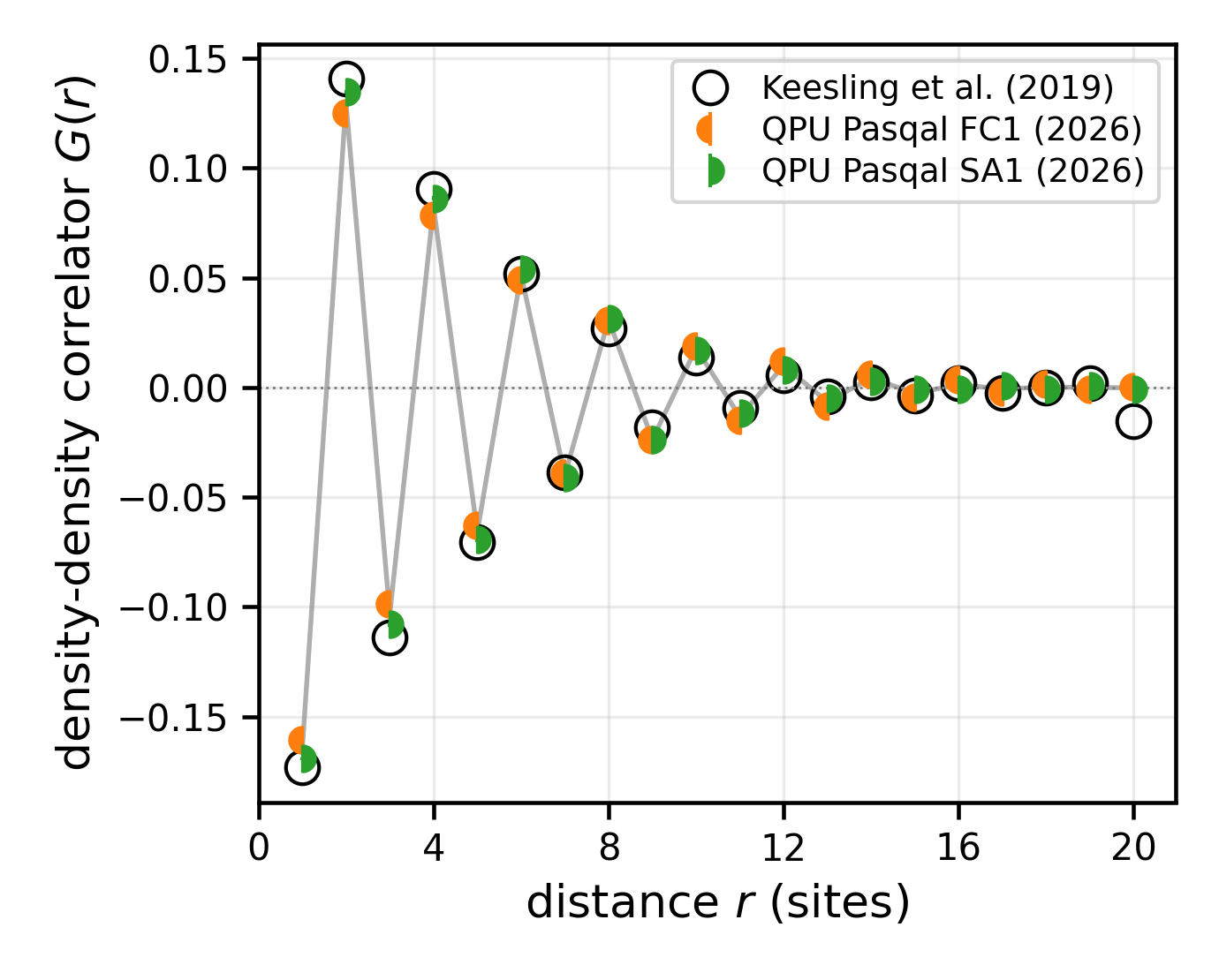}
    \caption{Density-density correlator $G(r)$ in the saturated $\mathbb{Z}_2$-ordered
    phase as a function of the distance $r$ in number of sites. Open black circles: values taken from Fig.~1d of Keesling \textit{et al.}~\cite{keesling2019quantum}
    (51-atom chain, fitted $\xi=3.9$ sites). Half-filled markers: this work, on a
    56-atom ring on Pasqal QPU FC1 (orange, 2000 shots, $\xi=4.2$) and a
    96-atom ring on Pasqal QPU SA1 (green, 1000 shots, $\xi=4.2$). The grey line is a guide to the eye. All three datasets show that
    the saturated ordered-phase correlator is a robust observable, reproduced by our agentic workflow on two Pasqal QPUs, 7 years after the landmark experiment.
    }
    \label{fig:keesling}
\end{figure}

These observations fit very well with our intuition, confirming that the agent identifies correctly the hardware constraints.

Redirected to the $\mathbb{Z}_2$ phase, the agent uses the $\texttt{noise_emulate}$ skill and obtains 80\,\% noise retention with the saturated order parameter, reaching 71\,\% of its ideal value. The order parameter is the density-density correlator introduced in the original work~\cite{keesling2019quantum}
\begin{equation}
    G(r) = \frac{1}{N_r}\sum_{i}\Big(\langle n_i n_{i+r}\rangle - \langle n_i\rangle\langle n_{i+r}\rangle\Big).
    \label{eq:Gr}
\end{equation} 

Prompted to make better use of the field of view, we suggest replacing the open chain with a ring register to add atoms and remove edge effects. The agent then sizes a 56-atom ring at the spacing floor and reruns the pipeline. We furthermore note that the second Pasqal QPU SA1 offers a larger field of view. The agent sizes a 96-atom ring for it.

In Fig.~\ref{fig:keesling}, we show that the reconstructed correlator from the 2019 apparatus and from both 2026 devices coincide to within statistical error at all 20 lattice sites probed, with fitted correlation lengths of $\xi=3.9$ (paper), $4.2$ (FC1) and $4.2$ (SA1) sites.

\subsubsection{Comments}

This first example already demonstrates excellent agentic capability. In addition to independently constructing the noiseless-vs-ideal comparison, it also recovered from infrastructure failures (a paused QPU, errors in jobs) without derailing the broader campaign. Agents have the healthy habit of consistently adding checkpoints, error logs and try/except errors that help recover from failures. Most noticeably, the pipeline was extremely fast: in less than an hour and three researcher prompts the agentic workflow had QPU jobs up and running.

That said, the higher-level scientific moves that improved the outcome, such as switching to a ring register or extending the study to a second QPU originated from the researcher. Note that every such piece of input is retained by the agentic workflow, so that future experiments start from an accumulated body of domain know-how rather than repeating the same reasoning from scratch. Once directed, the agent executed each multi-hour, multi-device campaign autonomously.

\subsection{Preparation of frustrated quantum magnetic phases on a triangular lattice}\label{sec:ex_guo}

We now turn to a harder use-case for which a theory paper exists but no QPU implementation is proposed.

\subsubsection{Context and task}

Guo \emph{et al.}~\cite{guo2023triangular} explore the quantum magnetic phases of a
frustrated triangular Rydberg-atom array. Using quantum Monte Carlo, the authors map
out a phase diagram containing three ordered lobes as a function of the reduced
detuning $\delta/\Omega$: a $1/3$-filling plateau, a
$1/2$-filling ``order-by-disorder'' phase known to be experimentally fragile, and a
$2/3$-filling plateau.

We give the agent a deliberately open-ended system prompt. This tests whether it can autonomously isolate the relevant physical variables and map out an experimental realisation:

\begin{promptbox}[title={Agent Prompt: Phase Evaluation}]
Read Guo et al. paper (link provided). For each distinct phase of matter
described in the paper, assess whether it can realistically be prepared and observed
on Pasqal QPU FC1 using our pipeline. Produce a verdict table (feasible /
marginal / infeasible) with your reasoning. For every phase you judge feasible or
marginal, output a complete experiment\_spec.json ready for spec-to-sequence.
\end{promptbox}

From an expert perspective, preparing a frustrated ground state on hardware requires
several trade-offs: choosing commensurate system sizes, designing an adiabatic
annealing trajectory, and tracking a combination of observables sufficient to certify
the phase unambiguously. Our goal is to assess how effectively the agent navigates
these nuances.

\subsubsection{Workflow and dialogue}

Invoking the \texttt{paper-to-spec} skill, the agent correctly parsed the paper's core
results, identifying the $1/3$ and $2/3$ plateaus alongside the delicate $1/2$-filling
order-by-disorder phase. Its verdict table again demonstrated a firm grasp of the
hardware constraints. The decisive limitation is FC1's maximum detuning,
$|\delta|/2\pi \le 10$~MHz: since each lobe is specified by a fixed reduced detuning
$\delta/\Omega$, reaching a deeper lobe forces a proportionally smaller Rabi
frequency, and hence a longer adiabatic ramp measured against a fixed coherence time.
Working from the tunable knobs the agent reasoned that

\begin{itemize}
    \item[\textit{(i)}] the $1/3$-filling phase ($\delta_{c1}/\Omega = 1.22$) is
    \emph{feasible}, since its lobe is reached at the full
    $\Omega/2\pi = 2$~MHz;
    \item[\textit{(ii)}] the $1/2$-filling phase ($\delta/\Omega \simeq 9.5$) is only
    \emph{marginally} feasible, as it requires halving the Rabi frequency to
    $\Omega/2\pi \simeq 1$~MHz, which lengthens the ramp and risks non-adiabatic
    transitions out of the ground state;
    \item[\textit{(iii)}] the $2/3$-filling phase ($\delta_{c2}/\Omega = 17.8$) is
    \emph{infeasible}, as it would demand $\Omega/2\pi \lesssim 0.6$~MHz, a ramp far
    too slow compared with the atomic coherence time.
\end{itemize}

\begin{figure}
\centering
\includegraphics[width=\linewidth]{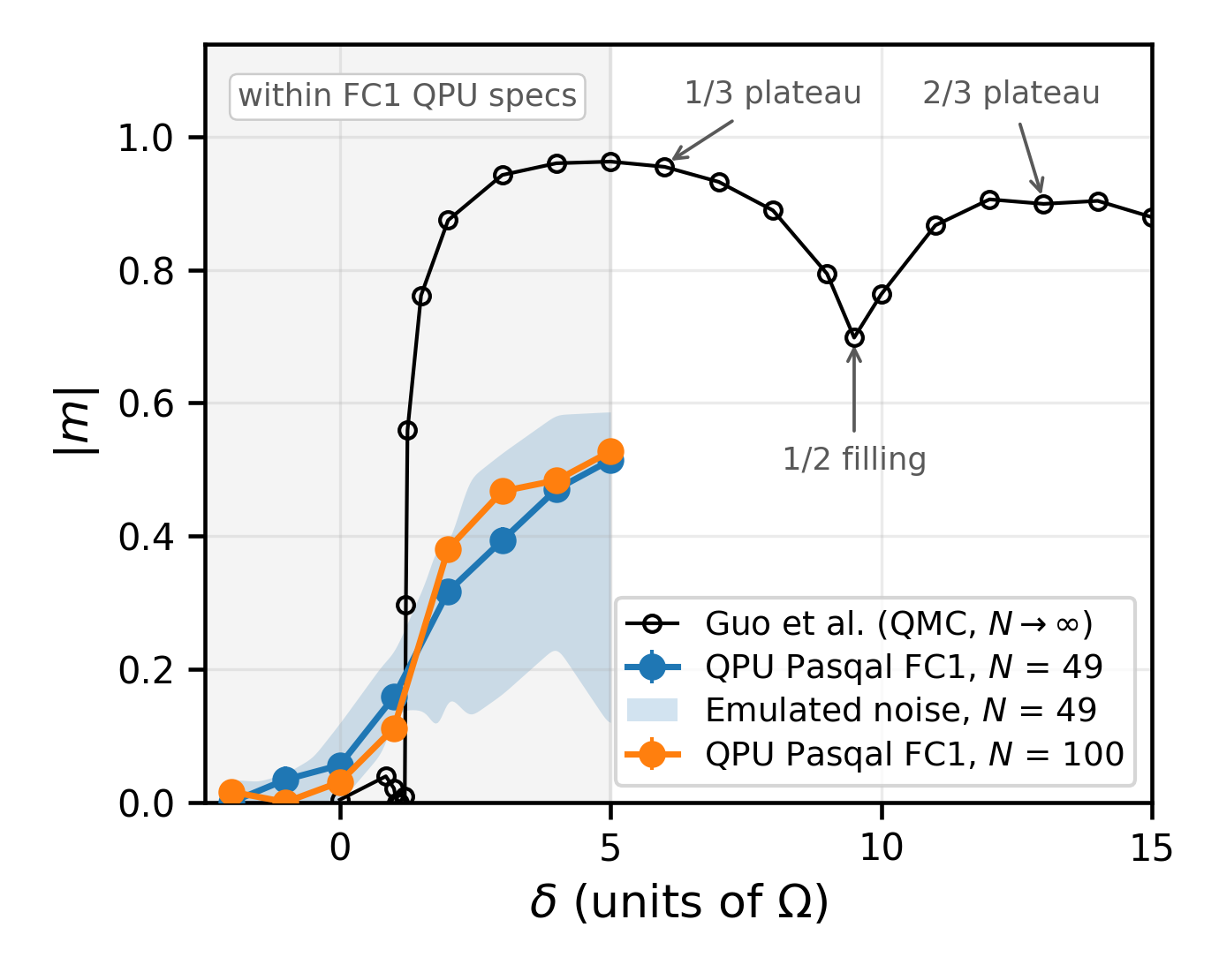}
\caption{Normalised order parameter $|m|$ versus $\delta / \Omega $ on the bulk sublattice. Open black circles: QMC in the thermodynamic limit (Guo et al.~\cite{guo2023triangular}). Filled circles: QPU Pasqal FC1 at $N=49$ (blue) and $N=100$
(orange), on rhombic $L\times L$ triangular registers with $L\equiv 1 \bmod 3$, so
the open boundary pins a single $\sqrt{3}$ sublattice.
Shaded band: MPS emulation of the $N=49$ register including the calibrated FC1
noise model ($\pm2\sigma$ over $40$ quantum-jump trajectories).
All data share $a=5.3\,\mu$m ($R_b/a=1.21$), $\Omega/2\pi=2$\,MHz, and a $4\,\mu$s
linear detuning ramp from $\delta_i=-4$\,MHz, with $300$ shots per point.}
\label{fig:guo_fig2a}
\end{figure}

We instructed the agent to generate and test \texttt{Pulser} sequences preparing the viable phases. Left to its own devices, the agent exhibited a form of cognitive narrowing: of the four
observables detailed in the original paper it retained only the structure factor and the mean magnetisation. While the structure factor is an appropriate two-body diagnostic, the mean magnetisation is not the one-body order parameter of the
paper, and cannot distinguish the particular pattern from any other state at the same density. The agent also overlooked the finite-size scaling arguments that are central to interpreting small registers. Guided by the researcher on both points, it identified the staggered magnetisation
\begin{equation}
    m \;=\; \frac{3}{N}\sum_{j=1}^{N} n_j \, e^{i\mathbf{Q}\cdot\mathbf{r}_j},
    \qquad
    \mathbf{Q} \;=\; \left(\frac{2\pi}{3a},\, \frac{2\pi}{\sqrt{3}\,a}\right),
    \label{eq:staggered_mag}
\end{equation}
where $n_j \in \{0,1\}$ is the Rydberg occupation of atom $j$ and the prefactor
normalises perfect $1/3$ order to $|m| = 1$. Because the phase is three-fold degenerate, the sublattice selected varies from shot to shot and $\langle m \rangle$ averages to zero, hence the importance of selecting the shot-resolved magnitude $\langle |m| \rangle$.
We then directed the agent to design a scaling study for commensurate rhombic registers ($L \times L$ with $L \equiv 1 \bmod 3$), yielding valid sizes $N=49$ ($L=7$) and $N=100$ ($L=10$). The resulting QPU data (Fig.~\ref{fig:guo_fig2a}) clearly display the $1/3$ plateau. A deficit of roughly half the QMC amplitude arises from finite-size rounding, diabatic ramp excitations, and hardware noise. The emulated noise faithfully envelopes the QPU datapoints. 

\subsubsection{Comments}

The agent demonstrated a significant ability to extract, translate, and execute a hardware experiment directly from a theory paper, eliminating tedious boilerplate script generation and hardware parameter mapping to sharply reduce the time-to-QPU.

However, we observed a key failure mode: the agent consistently optimized for the nearest tractable sub-task at the expense of the core physics, selecting the observable that was easiest to compute rather than the one that genuinely certifies the phase. Crucially, because the pipeline ran smoothly and generated plausible-looking data, only domain expertise could identify that the wrong physical quantity was being measured. Consequently, while Example~A required just 3 human messages, this example required 43 exchanges to reach a physically accurate implementation. This highlights that agentic workflows must be evaluated not merely on execution success, but on physical accuracy.

Finally, these experiments clarify current QPU boundaries. The $1/2$- and $2/3$-filling lobes are restricted here solely by the maximum detuning $|\delta|_{\max}$, and could be unlocked either by expanding the detuning range or rescaling the blockade ratio to shift the lobes to smaller $\delta/\Omega$, a regime already accessed on academic setups~\cite{scholl2021quantum}.

\begin{figure*}
\centering
\includegraphics[width=\textwidth]{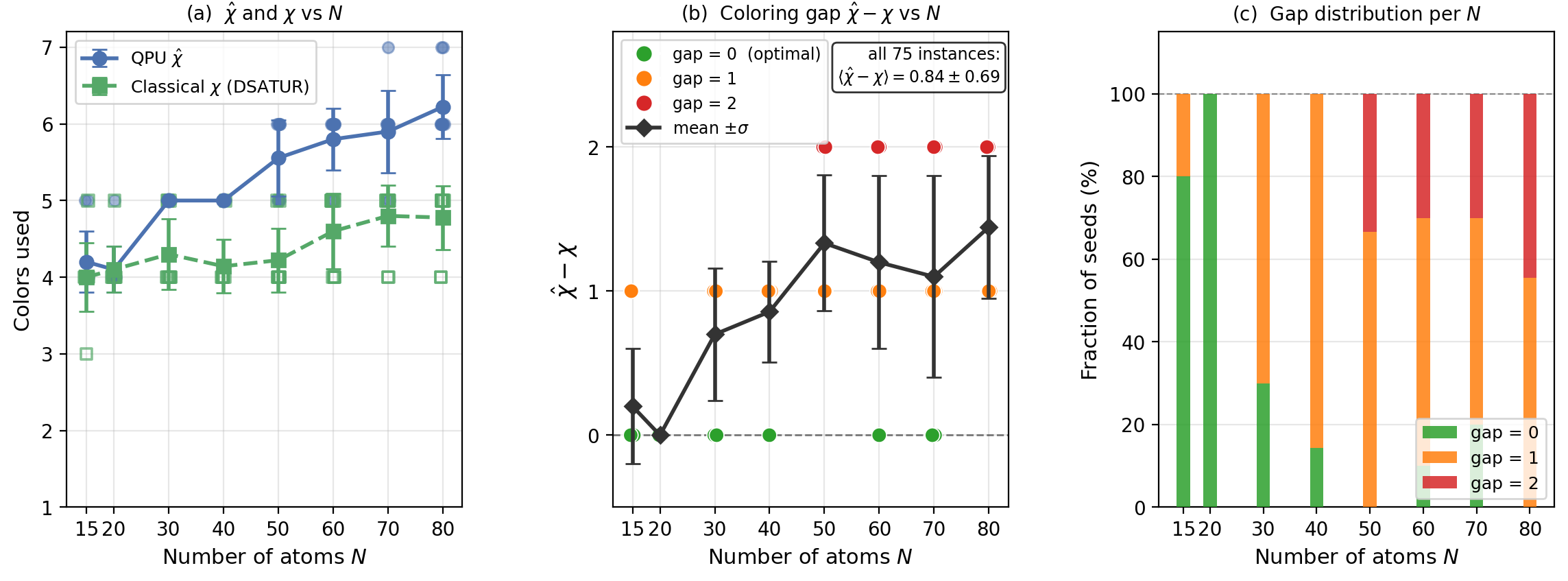}
\caption{Example~C, QPU beam-search graph coloring on FC1 (unit-disk
graphs at density $0.35$, blockade radius $R_b{=}10\,\mu$m; beam-width
$w{=}3$, $n_s{=}200$ shots/round, 75~instances).
(a)~QPU estimate $\hat\chi$ and classical optimum $\chi$ (DSATUR) vs.\ $N$;
lines show mean~$\pm\sigma$, dots show per-instance values.
(b)~Coloring gap $\hat\chi{-}\chi$ per instance (coloured by gap value) and
mean~$\pm\sigma$; the gap is zero at $N{=}20$ and never exceeds~2.
(c)~Fraction of seeds at each gap value per $N$.}
\label{fig:coloring_scaling}
\end{figure*}

\subsection{Patent implementation: graph coloring via iterative Rydberg IS peeling}\label{sec:ex_coloring}

We now move to our last example, taken from combinatorial optimization. The source here is distinctive: not a journal paper but a Pasqal \emph{patent}. This
tests whether the same workflow mines patents as readily as the published literature, a second and largely untapped corpus of hardware-ready protocols. The patent specifies the algorithm but not its realization on present hardware, so the agent must bridge from a protocol description to a working multi-round pipeline.
\subsubsection{Context and task}

Graph coloring (assigning colors to the vertices of $G=(V,E)$ such that adjacent
vertices receive different colors) is equivalent to a sequence of independent-set (IS)
problems: peel one IS per round until all vertices are covered, with the round count
upper-bounding the chromatic number $\chi(G)$. A \emph{project-and-expel} algorithm maps
this decomposition onto the native Rydberg IS
primitive~\cite{henriet2026coloring,vercellino2023bbqmis}: starting from all $N$ atoms in
$|0\rangle$, a resonant global pulse drives the current sub-register into a superposition
of independent sets; one measurement shot collapses onto a valid IS; those atoms are
assigned a color and retired; the residual subgraph defines the next round.

\begin{promptbox}[title={Agent Prompt}]
Read the attached Pasqal patent describing the project-and-expel algorithm for
estimating the chromatic number of a graph on a Rydberg array. Implement it on
Pasqal QPU FC1: map the iterative IS-peeling decomposition onto the native Rydberg IS
primitive, handle the multi-round orchestration on present hardware, and benchmark the
estimated chromatic number against a classical baseline on random unit-disk graphs.
Document any hardware constraint that forces a deviation from the patent's idealized
procedure.
\end{promptbox}

\subsubsection{Workflow and dialogue}

On present hardware, physical atom expulsion is replaced by a classical
measurement-feedback loop: the agent compiles the full $N$-atom layout once into a
\texttt{MappableRegister} batch and submits each round as a separate job within that
open batch, activating only the subset of trap sites corresponding to the current
subgraph.
This eliminates re-queuing latency between rounds while holding a single QPU queue slot.
To cope with finite shot counts, the agent runs a beam search (width $w{=}3$,
$n_s{=}200$ shots per round per branch), maintaining three candidate colorings in
parallel and retaining the best final result.
A classical repair step corrects any blockade-imperfect shots that violate the IS
constraint.

The multi-round orchestration rests on Pulser~\cite{silverio2022pulser} for
pulse-level compilation and the Pasqal Cloud SDK for batch lifecycle management.
The use of \texttt{MappableRegister} and open-batch primitives was not specified
a priori; the agent designed and refined this pipeline through iterative
back-and-forth with the user, consulting the SDK and hardware documentation to
identify constraints that do not surface programmatically---notably, the minimum
trap-site filling fraction the cloud API enforces when the active subgraph shrinks
in late rounds, which would otherwise cause job rejections.

We benchmark on random unit-disk graphs with
$N\in\{15,20,30,40,50,60,70,80\}$ atoms, 75~instances in total.
The classical chromatic number $\chi$ (DSATUR) is 4--5 throughout, confirming that
$\chi$ stays bounded as $N$ grows from 15 to 80 for these sparse graphs.
At $N{=}20$ the QPU achieves the exact chromatic number on all 10~instances
(mean gap $\langle\hat\chi-\chi\rangle{=}0.00$); the gap grows to at most
$1.44\pm0.50$ at $N{=}80$, and never exceeds~2 across all 75~instances
(overall mean $0.84$).
Figure~\ref{fig:coloring_scaling} shows the scaling.

\subsubsection{Comments}

This example extends the workflow beyond a single-shot protocol to a
\emph{multi-round, state-dependent} pipeline: the agent compiles the register once,
routes $\hat\chi$ sequential jobs within one open batch, manages beam-search branches
across rounds, and aggregates the final coloring, an orchestration pattern that would
require substantial bespoke engineering outside an agentic scaffold.
It also demonstrates that the corpus an agent can draw on is not limited to published
papers: a patent, read directly, served as a complete and actionable protocol source. 

As in the previous example, though, the agent required direction from the researcher and
occasionally went astray. Confronted with an unexpectedly high rate of
constraint-violating shots, the agent confidently attributed it to physics, diagnosing
substantial Rydberg-blockade leakage on the QPU. The true cause was a mismatch
between the ordering of the atomic coordinates and the ordering of the measured
bitstrings, which scrambled the vertex labels and made valid independent sets appear to
violate the adjacency constraints. The pathology lay in the workflow, neither the hardware nor the physics, and only expert scrutiny of the raw shot-to-graph mapping surfaced the labelling bug behind the agent's plausible but incorrect physical story. 

\section{Corpus-scale analysis of the Rydberg-array literature}\label{sec:all_rydberg_lit}

The examples in the previous section demonstrate the workflow on individual protocols. The same agentic approach can also be applied at corpus scale: a second agentic workflow surveys the Rydberg-array theory literature and performs a rapid feasibility assessment based on the physical resources required by each proposal. The complete corpus-construction procedure, search queries, filtering pipeline, and classification workflow are described in Appendix~\ref{app:queries}.

The search identified 633 Rydberg-array theory papers. Of these, 107 could not be reliably classified because the corresponding PDF was unreadable or insufficiently detailed for automated analysis. The remaining 526 papers were classified according to the hardware capabilities required to implement their proposed protocols. Table~\ref{tab:corpus} summarizes the resulting distribution.

Among the classified papers, 258 ($49.0\%$) were found to be compatible with publicly available Pasqal QPUs. Of these, 225 describe sufficiently complete experimental protocols that can be translated directly into executable implementations, while the remaining 33 require modest adaptation, such as specifying an observable, pulse schedule, or parameter range. The remaining 268 papers require hardware capabilities that are not yet publicly available, most commonly XY-type interactions (141 papers) or local addressing (87 papers).

The 258 implementable papers represent viable starting points for experimentation on present-day hardware. In practice, obtaining a meaningful experimental signal still requires adapting each protocol to the characteristics of the device and selecting observables that remain robust in the presence of hardware noise. The agent substantially reduces this engineering effort while scientific judgment remains essential when designing the experiment and interpreting its outcome.

\begin{table}[t]
  \centering
  \caption{Classification of the 526 Rydberg-array theory papers the
    agent could read and classify (out of 633 on-topic; 107
    unclassified). The top two rows (258 papers, $49.0\,\%$) are
    implementable on a present-day neutral-atom QPU; the remaining rows
    group the rest by which rule they fail.}
  \label{tab:corpus}
  \begin{tabular}{lrr}
    \toprule
    Class & Papers & \% \\
    \midrule
    Implementable today                              & 225 & 42.8 \\
    Implementable with modest adaptation             &  33 &  6.3 \\
    \midrule
    Needs different atomic encoding (XY)             & 141 & 26.8 \\
    Needs local addressing                           &  87 & 16.5 \\
    Too big ($N > 100$)                              &  17 &  3.2 \\
    Needs measurements during the pulse              &   9 &  1.7 \\
    Needs basis rotations beyond on/off readout      &   5 &  1.0 \\
    Genuinely under-specified                        &   9 &  1.7 \\
    \midrule
    Total                                            & 526 & 100  \\
    \bottomrule
  \end{tabular}
\end{table}

Beyond estimating the current experimental reach of neutral-atom platforms, the classification also provides a quantitative view of future hardware priorities. The vast majority of currently inaccessible protocols require either XY interactions or local addressing, which together account for approximately $85\%$ of the non-implementable papers. The same analysis therefore serves two complementary purposes: it estimates how much of today's Rydberg-array literature can already be brought to hardware using an agentic workflow, and it identifies the hardware capabilities that would unlock the largest fraction of the remaining scientific literature.

\section{Summary and outlook}\label{sec:ccl}

While quantum hardware has become increasingly accessible through cloud platforms and high-performance computing infrastructures, translating a scientific idea into a hardware experiment remains a substantial undertaking. In this work, we introduced an agentic framework designed to bridge this gap, structured as a collection of modular AI-agent skills. By operating on a common experimental specification and state representation, these skills can be iteratively orchestrated by the researcher into refinement loops that regenerate sequences, run simulations, or adjust parameters to accelerate the time to run a real experiment on a Pasqal QPU.

Beyond these individual demonstrations, our corpus-scale analysis provides a broader perspective: approximately half of the analyzed theory papers describe protocols that are compatible with Pasqal QPUs and could immediately be explored experimentally using our workflow. Concurrently, the analysis of the remaining papers quantifies exactly which hardware capabilities—notably XY interactions and local addressing—would unlock the largest fraction of the literature, providing a demand-driven perspective for future hardware development.

A central observation emerging from our study is that the operational bottleneck has shifted. The agent proves highly effective at coordinating software tools, navigating device interfaces and generating experiments on the QPU. Its limitations appear rather in the selection of physically meaningful observables, recognizing subtle scientific assumptions, and deciding whether an experimental signal genuinely supports a physical claim. If anything, the researcher's role becomes more decisive than ever---hypothesis formulation, critical evaluation, and scientific interpretation of the data remain irreducibly human---now that the surrounding engineering no longer competes for their attention. The same tooling works in the theorist's favour upstream as well: by pairing noiseless predictions with noise-aware emulation, it gives a rapid, quantitative read on which observables are likely to rise above the hardware noise, concentrating effort on the signals a QPU can actually resolve.

Looking forward, we believe this work also paves the way for scientific reproducibility. Reproducing a published quantum experiment often requires reconstructing a large amount of implicit engineering knowledge that is only partially captured in the manuscript itself. By translating scientific protocols into structured, executable specifications, agentic workflows offer a scalable framework to preserve this engineering layer alongside the publication. As hardware and software platforms evolve, these specifications can be re-instantiated, adapted to new devices, and revalidated with minimal manual effort. In this sense, publications can evolve from static descriptions of experiments into executable scientific artifacts that can be systematically revisited, verified, and extended over time.

Ultimately, we expect that the most transformative quantum applications will emerge when quantum hardware becomes accessible not only to the teams that build it, but also to a broader scientific community capable of bringing new questions, unique datasets, and fresh perspectives to the field~\cite{jurczak2026qoulipo}.

\bibliography{references}

\vspace*{\baselineskip}

\section*{Acknowledgments}
The hardware results reported here were obtained on the Pasqal QPUs FC1 and SA1 through standard Pasqal Cloud accounts available to external users. FC1 is located in Sherbrooke, in the DistriQ quantum-innovation zone (Québec, Canada); SA1 is located in the Kingdom of Saudi Arabia. The AI coding agent used throughout this work is Anthropic's Claude Fable 5 and Opus 4.8.
\appendix

\section{Agentic workflow description}
\label{app:workflow}
The framework exposes six modular skills, which share a common per-experiment state and exchange information through structured artifacts.

\begin{itemize}
    \item $\texttt{paper_to_spec}$ translates a scientific objective into a structured experimental specification. The input may be a research publication, patent, technical report, or a high-level user description. The resulting $\texttt{experiment_spec.json}$ captures the relevant physical parameters, observables, and experimental constraints. \\ 
    \item $\texttt{spec_to_sequence}$ converts the specification into an executable Pulser implementation. In addition to generating the pulse sequence itself, the skill exposes a standard $\texttt{build_sequence}$ interface that is reused by the simulation, validation, calibration, and hardware-execution stages, so that modifications to the specification propagate consistently downstream. \\
    \item $\texttt{validate_emu}$ evaluates a candidate protocol with $\texttt{emu-mps}$. A typical validation pairs a noiseless simulation, which provides the reference prediction, with a noise-aware simulation that incorporates device-specific noise parameters obtained through the Pasqal Cloud SDK and generates a statistical envelope from multiple quantum-jump trajectories. This skill is aimed in particular at Pulser-literate users who may not be aware of the more delicate subtleties of quantum hardware. \\
    \item $\texttt{noise_emulate}$ performs a high-fidelity noisy emulation of a candidate sequence on local GPU resources, complementing the faster cloud-based screening of $\texttt{validate_emu}$. The sequence is simulated with matrix-product-state trajectories: a single noiseless reference run plus an ensemble of forty noisy trajectories, whose statistics define a quantile envelope for each observable. The underlying noise model is fetched live from the device through the Pasqal Cloud SDK at execution time, so that the emulation reflects the current operating point of the hardware rather than static datasheet values. Finally, the envelope is widened by a calibration-offset band that propagates residual uncertainty in the Rabi frequency and detuning calibration, producing the acceptance region against which measured QPU data are later evaluated by $\texttt{harvest_and_analyze}$. \\

    \item $\texttt{qpu_submit}$ manages execution on quantum hardware. Before submission, it can run a short calibration procedure to characterize the current operating point of the device; the resulting calibration values are propagated to both the experimental parameters and the corresponding noise models. The calibrated experiment is then submitted through the Pasqal Cloud SDK. \\ 
    \item $\texttt{harvest_and_analyze}$ retrieves completed experimental data and produces higher-level scientific observables. Raw bitstrings are processed using device-reported corrections and compared against both the noiseless prediction and the noise-aware simulation envelope, giving a direct assessment of the agreement between theory, hardware-aware simulation, and measurement.
\end{itemize}

\section{Details of the search query}\label{app:queries}

We issued nine topical queries against the arXiv abstract index, retrieving 1{,}966 raw hits, which deduplicated to 1{,}108 unique papers. A subsequent filtering stage removed off-topic matches (e.g.\ Fresnel optics, Pulser signal-processing literature, and Rydberg vapour-cell sensor work), resulting in a final corpus of 633 Rydberg-array theory papers.

Each paper was then processed by an automated agent that parses the full text PDF and assigns a protocol class using a five-rule decision tree:

(i) system size $N \le 100$;  
(ii) global drive only (i.e.\ no site-selective driving or local detuning);  
(iii) Ising or blockade interactions (excluding XY models, resonant dipole--dipole exchange, dual-species architectures, and microwave-dressed couplings);  
(iv) absence of mid-circuit measurement; and  
(v) computational-basis bitstring readout without basis rotations for tomography.

Papers satisfying all five criteria are marked as \emph{implementable}. Otherwise, the first violated rule determines a hardware feature label indicating the missing capability (see Table~\ref{tab:corpus}). For each classification, the agent stores rule-level justification with citations to the relevant section or figure in the paper, enabling auditability without requiring full manual review.

\begin{table}[t]
  \centering
  \caption{Verbatim arXiv \texttt{search\_query} strings used in the corpus scan (executed 2026-05-23) and corresponding hit counts per query. All queries target the abstract field via the \texttt{abs:} prefix. The aggregate count includes overlaps across queries; deduplication yields 1{,}108 unique papers.}
  \small
  \begin{tabular}{lp{0.55\linewidth}r}
    \toprule
    Label & \texttt{search\_query} & Hits \\
    \midrule
    \texttt{rydberg\_array\_dyn} &
      \texttt{abs:"rydberg atom array" AND (abs:dynamics OR abs:quench OR abs:sweep)} &
      106 \\[2pt]

    \texttt{rydberg\_scars} &
      \texttt{abs:"rydberg" AND (abs:scar OR abs:"many-body scar" OR abs:"Z2 state" OR abs:"Z3 state")} &
      100 \\[2pt]

    \texttt{rydberg\_mis\_opt} &
      \texttt{abs:"rydberg" AND (abs:"maximum independent set" OR abs:"MIS" OR abs:"combinatorial optimization")} &
      49 \\[2pt]

    \texttt{rydberg\_floquet} &
      \texttt{abs:"rydberg" AND (abs:"Floquet" OR abs:"periodic driv")} &
      91 \\[2pt]

    \texttt{rydberg\_phase\_trans} &
      \texttt{abs:"rydberg atom array" AND (abs:"phase transition" OR abs:"ordered phase" OR abs:"adiabatic")} &
      68 \\[2pt]

    \texttt{rydberg\_2d\_lattices} &
      \texttt{abs:"rydberg" AND (abs:kagome OR abs:triangular OR abs:"square lattice")} &
      115 \\[2pt]

    \texttt{pulser\_pasqal\_sp} &
      \texttt{abs:"Pulser" OR abs:"Pasqal" OR abs:"FRESNEL"} &
      1{,}188 \\[2pt]

    \texttt{rydberg\_xy\_dipolar} &
      \texttt{abs:"rydberg" AND (abs:"XY model" OR abs:"dipolar" OR abs:"resonant dipole" OR abs:"flip-flop")} &
      204 \\[2pt]

    \texttt{rydberg\_topology} &
      \texttt{abs:"rydberg atom array" AND (abs:topological OR abs:"spin liquid" OR abs:"gauge theory")} &
      45 \\
    \midrule
    Sum (with overlap)         & & 1{,}966 \\
    Unique & & 1{,}108 \\
    \bottomrule
  \end{tabular}
\end{table}

\end{document}